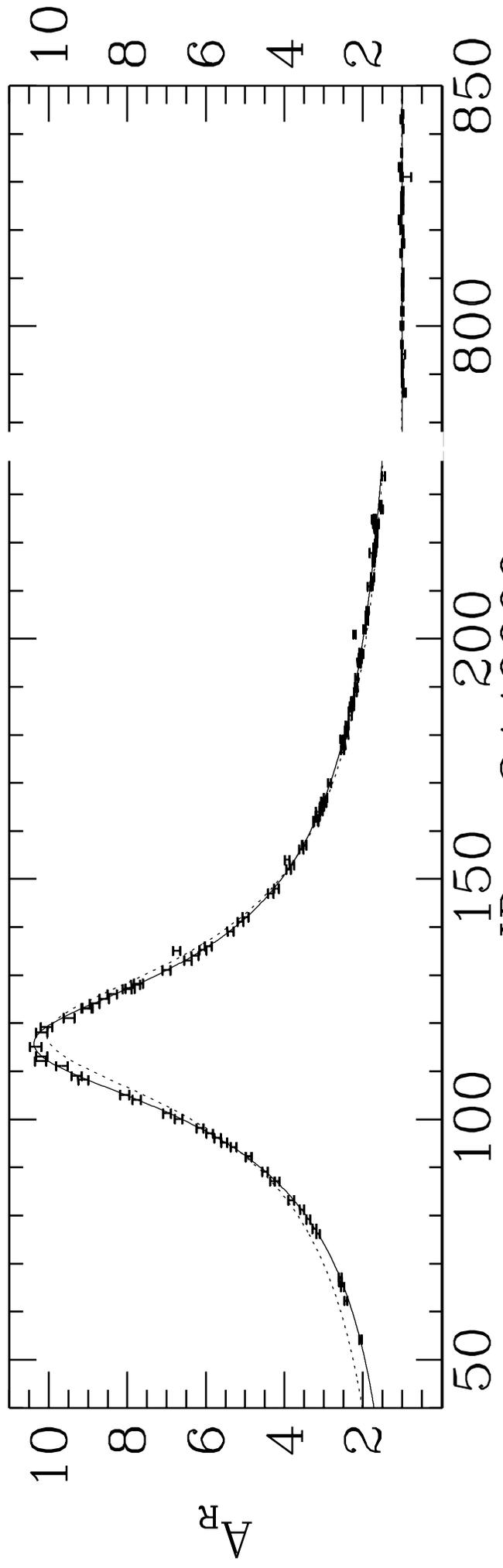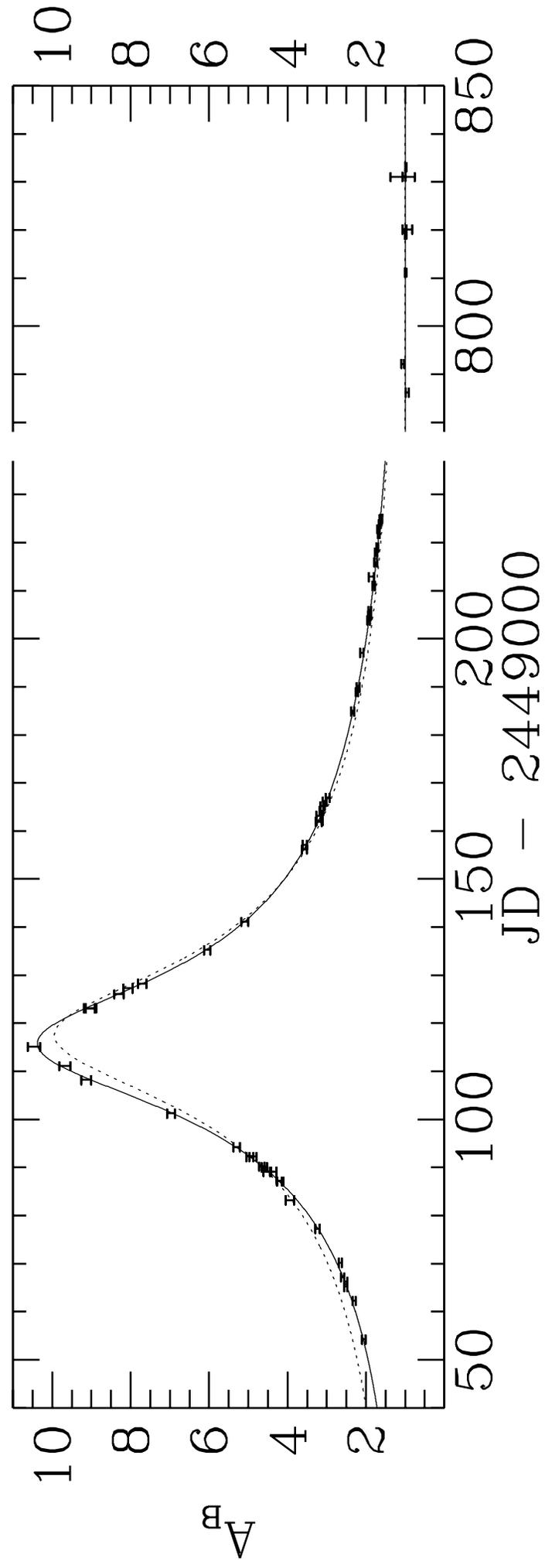

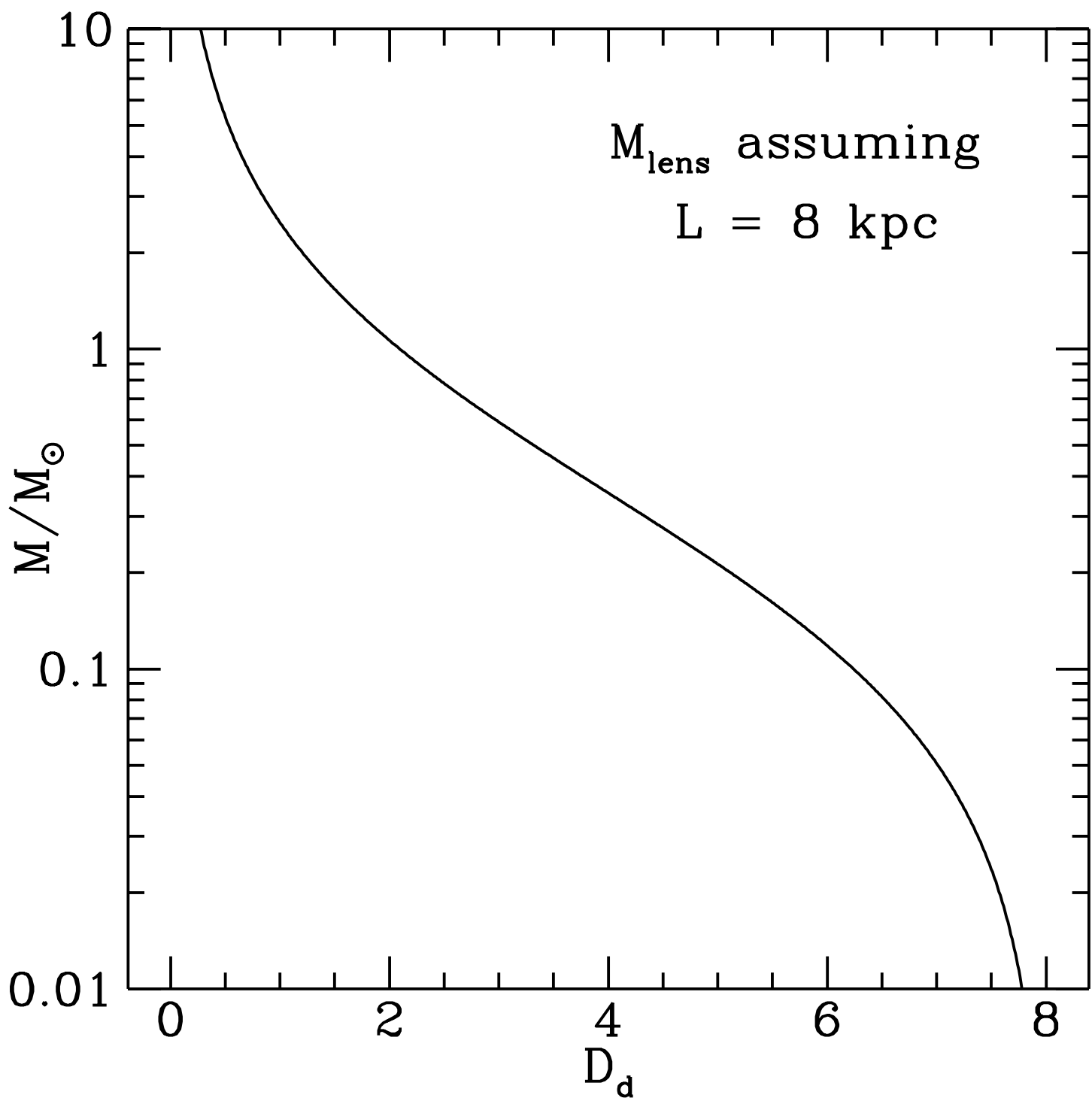

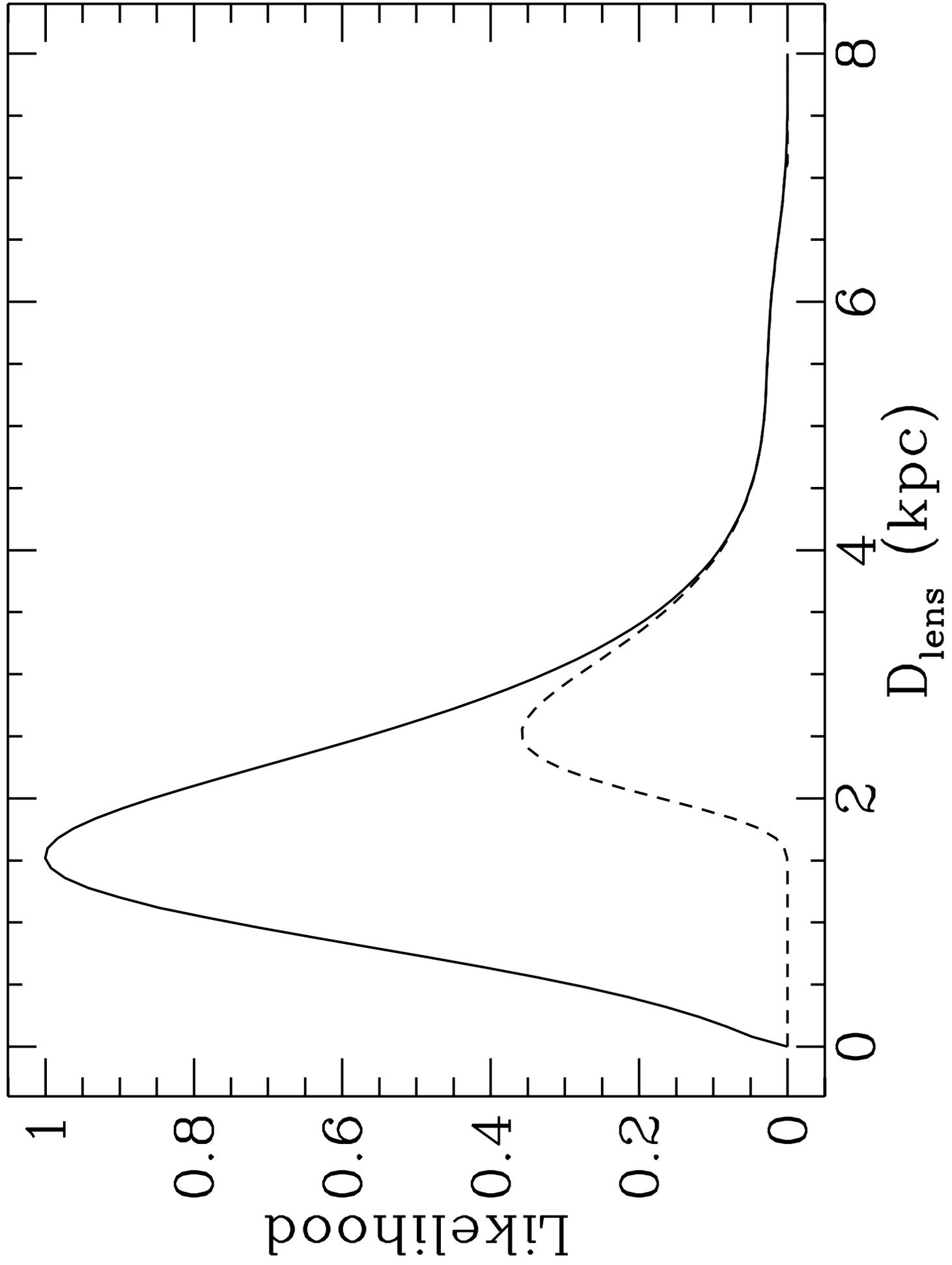

# First Observation of Parallax in a Gravitational Microlensing Event

C. Alcock[1,2], R.A. Allsman[3], D. Alves[1,4], T.S. Axelrod[1,5],

D.P. Bennett[1,2], K.H. Cook[1,2], K.C. Freeman[4], K. Griest[2,6],

J. Guern[2,6], M.J. Lehner[2,6], S.L. Marshall[2,7],

B.A. Peterson[5], M.R. Pratt[2,7], P.J. Quinn[5], A.W. Rodgers[5]

C.W. Stubbs[2,7,8], W. Sutherland[2,9]

(The MACHO Collaboration)

1 : Lawrence Livermore National Laboratory, Livermore, CA 94550

2 : Center for Particle Astrophysics,

University of California, Berkeley, CA 94720

3 : Supercomputing Facility,

Australian National University, Canberra, A.C.T. 0200, Australia

4 : Department of Physics, University of California, Davis, CA 95616

5 : Mt. Stromlo and Siding Spring Observatories,

Australian National University, Weston, A.C.T. 2611, Australia

6 : Department of Physics, University of California, San Diego, CA 92039

7 : Department of Physics, University of California, Santa Barbara, CA 93106

8 : Departments of Astronomy and Physics,

University of Washington, Seattle, WA 98195

9 : Department of Physics, University of Oxford, Oxford OX1 3RH, U.K.

# ABSTRACT

We present the first detection of parallax effects in a gravitational microlensing event. Parallax in a gravitational microlensing event observed only from the Earth appears as a distortion of the (otherwise symmetrical) lightcurve due to the motion of the Earth around the Sun. This distortion can be detected if the event duration is not much less than a year and if the projected velocity of the lens is not much larger than the orbital velocity of the Earth about the Sun. The event presented here has a duration (or Einstein diameter crossing time) of $\hat{t} = 220$ days and clearly shows the distortion due to the Earth's motion. We find that the projected velocity of the lens is $\tilde{v} = 75 \pm 5$ km/s at an angle of $28° \pm 4$ from the direction of increasing galactic longitude, as expected for a lens in the galactic disk.

A likelihood analysis of this event yields estimates of the distance to and mass of the lens: $D_{\text{lens}} = 1.7^{+1.1}_{-0.7}$ kpc and $M = 1.3^{+1.3}_{-0.6} M_\odot$. This suggests that the lens is a remnant such as a white dwarf or neutron star. It is possible, though less likely, that the lens is a main sequence star. If so, we can add our upper limit on the observed flux from the lens to the analysis. This modifies the estimates to: $D_{\text{lens}} = 2.8^{+1.1}_{-0.6}$ kpc, and $M_{\text{lens}} = 0.6^{+0.4}_{-0.2} M_\odot$.

## Introduction

In less than two years since the first candidate gravitational microlensing events were discovered (Alcock *et al.* 1993, Aubourg *et al.* 1993, Udalski *et al.* 1993), microlensing has been confirmed spectroscopically (Benetti *et al.* 1995) and more than 60 microlensing events have now been discovered (Udalski *et al.* 1994, Alcock *et al.* 1995a, 1995b, Bennett *et al.* 1995). The vast majority of the events have been discovered towards the Galactic bulge, and the total microlensing optical depth seen towards the bulge seems to be a factor of $\sim 3$ larger than predicted (Griest *et al.* 1992, Paczyński 1992). These results suggest that standard Galactic models are probably in need of revision (Paczyński *et al.* 1994, Zhou *et al.* 1994, Gould 1994), and this may have important implications for the interpretation of microlensing results toward the LMC (Alcock *et al.* 1995a, Aubourg *et al.* 1995).

Microlensing light curves are generally distinguished from other types of stellar variability by their very simple form. This is a great benefit when trying to detect these very rare events, but it makes their interpretation more difficult because a wide variety of lensing events can give rise to very similar light curves. In this paper, we present the first detection of the microlensing parallax effect which, as we shall see, offers great promise for resolving the degeneracies in microlensing events.

In most microlensing events, it is a very good approximation to assume that the source star is point-like, the lens mass is also point-like, and that the velocities of the source, the lens, and the observer are all constant in time. Under these assumptions, microlensing light curves take a very simple form: the amplification is given by

$$A(t) = \frac{u^2 + 2}{u\sqrt{u^2 + 4}} \; ; \quad u(t) \equiv \sqrt{u_0^2 + [2(t - t_0)/\hat{t})^2]} \tag{1}$$

where $\hat{t}$ is the Einstein diameter crossing time defined by

$$\hat{t}/2 = R_E/v_t \; ; \quad R_E^2 \equiv 4GMx(1-x)L/c^2 \; , \tag{2}$$

where $R_E$ is the Einstein ring radius, $v_t$ is the transverse velocity of the lens relative to the (moving) observer-source line, $M$ is the lens mass, $L$ is the observer-source distance, $x$ is the ratio of the observer-lens and observer-source distances, and $u_0 = b/R_E$ where $b$ is the closest approach of the lens to the observer-source line. We can see from eqs. (1) and (2) that $A(t)$ is described by only three parameters: $u_0$, $t_0$, and $\hat{t}$ (in practice, there is an additional parameter for the unlensed flux of the star in each passband). Furthermore, the parameters $u_0$ and $t_0$ reveal only the relatively uninteresting information about when the lens came closest to the line of sight and how close it came. The information about $M$, $x$, and $v_t$, which would help us determine whether the lensing object is part of a bulge or disk population, is all folded into the single parameter $\hat{t}$, so it is impossible to determine these parameters separately (though probability distributions may be obtained, e.g. Griest *et al.*1992).

The situation is improved for events in which one or more of the assumptions leading to the generic 3-parameter light curves described in eqs. (1) and (2) are violated, (e.g., Mao & Paczyński 1991; Gould 1992, 1994; Gould *et al.* 1995). The MACHO collaboration has observed two such events in the first year bulge data. One is the event presented in this paper and the other is the binary lensing event first seen by OGLE (Udalski *et al.* 1994b). Our data on the binary event resolves the finite size of the source star (Bennett *et al.* 1995), and when our analysis of this event is complete, it should yield two constraints on combinations of $M$, $x$, and the lens velocity.

The situation is similar for the event described here, which is the longest of 45 events in our first-year bulge data. The light curve for this event is shown in Fig. 1. The dashed curve is the best-fit symmetric light curve described by eqs. (1) and (2). Clearly, the light curve of this star exhibits a significant asymmetry and has a large deviation from the best-fit symmetric light curve. However, the light curve is still achromatic, so it is likely to be a microlensing event. Such an asymmetry could be caused by a deviation from the constant velocity assumption for the source, lens, or observer. In principle, it is impossible to strictly distinguish between these three possibilities because each can contribute equally to $\mathbf{v}_t$, which is the transverse velocity of the lens with respect to the line of sight to the source star. However, since the orbit of the Earth is known, one can attempt to fit the light curve under the assumption that the only significant deviation from constant velocities is the orbit of the Earth. If a good fit is obtained assuming the known period, orientation and phase of the Earth's orbit, then it is reasonable to assume that the orbit of the Earth is indeed responsible for the deviation of the light curve because it is quite unlikely that the orbital parameters of the source star or the lens would exactly match those of the Earth. As we shall see, the event presented in this paper is a case where the orbital motion of the Earth is probably responsible for the deviation of the lightcurve. This is a fortunate circumstance as it allows us to compare the projected Einstein ring diameter crossing time with the known size of the Earth's orbit and thereby obtain a second constraint on the three unknown parameters of the lens, $M$, $x$, and $v_t$.

The star shown in Fig. 1 is located at $\alpha = 18^h\,03^m\,34.05^s$, $\delta = -28°00'18.9''$ (J2000) which in ecliptic coordinates is $\lambda = 270.85°$ and $\beta = -4.65°$. The small value of $|\beta|$ is a potential difficulty for our attempt to solve for the mass and location of the lens because the sign of the perpendicular component of the projected transverse velocity becomes undetermined in the $\beta = 0$ limit. To include the orbital motion of the Earth in eq. (1), we must replace the expression for $u(t)$ with

$$\begin{aligned}u^2(t) =& u_0^2 + \omega^2(t-t_0)^2 + \alpha^2 \sin^2[\Omega(t-t_c)] \\ &+ 2\alpha \sin[\Omega(t-t_c)]\left[\omega(t-t_0)\sin\theta + u_0\cos\theta\right] \\ &+ \alpha^2 \sin^2\beta \cos^2[\Omega(t-t_c)] + 2\alpha \sin\beta \cos[\Omega(t-t_c)]\left[\omega(t-t_0)\cos\theta - u_0\sin\theta\right]\end{aligned} \quad (3)$$

where $\theta$ is the angle between $\mathbf{v}_t$ and the North ecliptic axis, $\omega = 2/\hat{t}$, and $t_c$ is the time when the Earth is closest to the Sun-source line. Note that $u_0$ is no longer the minimum distance between the lens and the observer-source axis as it was in the constant velocity case, but is the minimum distance between the lens and the Sun-source line. The parameters $\alpha$ and $\Omega$ are given by

$$\alpha = \frac{\omega(1\text{AU})}{\tilde{v}}\left(1 - \epsilon\cos[\Omega_0(t-t_p)]\right) , \tag{4}$$

and

$$\Omega(t-t_c) = \Omega_0(t-t_c) + 2\epsilon\sin[\Omega_0(t-t_p)] , \tag{5}$$

where $t_p$ is the time of perihelion, $\tilde{v} = v_t/(1-x)$ is the transverse speed of the Macho projected to the position of the Sun, $\Omega_0 = 2\pi \text{ yr}^{-1}$, and $\epsilon = 0.017$ is the eccentricity of the Earth's orbit.

The dashed curve in Fig. 1 is the best-fit constant velocity microlensing light curve, and the solid curve is the fit including the motion of the Earth. The fit parameters and $\chi^2$ values are listed in Table 1. As seen from Table 1, the $\chi^2$ per degree of freedom is reduced dramatically from 8.42 to 1.54 when the terrestrial orbit terms are included. Formally, a $\chi^2$ per d.o.f. of 1.54 does not indicate a good fit for 206 degrees of freedom, but this is not unusual in our data for stars of this magnitude. If the 2 red band outlier points at days 135 and 201 are (arbitrarily) removed from the fit, the $\chi^2$ per d.o.f. drops to 1.08. In fact, these two outlier measurements do have unusually large point-spread-function fit $\chi^2$ values suggesting suspect photometry, but we do not impose an automatic cut on the PSF $\chi^2$ value for stars that are as bright as this.

| fit # | $t_0$ | $\omega$ (yrs$^{-1}$) | $u_0$ | $A_{\max}$ | $f_{0R}$ | $f_{0B}$ | $\tilde{v}$ ($\frac{\text{km}}{\text{sec}}$) | $\theta$ | $\chi^2$ |
|---|---|---|---|---|---|---|---|---|---|
| 1 | 143.7(8) | 3.28(4) | 0.159(11) | 10.38 | 1.000(3) | 1.000(4) | 75(5) | $-1.01(7)$ | 318.2 |
| 2 | 129.0(6) | 2.59(1) | 0.101(5) | 9.93 | 1.012(3) | 1.001(4) | $\infty$ | 0.0 | 1752 |

**Table 1:** Fit results for 213 measurements in both passbands. Fit # 1 is the full fit including parallax effects. Fit # 2 is the usual "constant velocity" fit, which is sufficient to describe most microlensing light curves. The error estimates given are the maximum extent of the surface in parameter space, which has $\chi^2$ greater than the best-fit value by 1.

The direction of the best-fit velocity, $\theta = -1.01$ rad from ecliptic North, is 28° away from the direction of increasing galactic longitude. This is what we expect for a lens located in the galactic disk, where much of the velocity should be due to the disk rotation, and thus the lens is 'overtaking' the line connecting the moving Sun to the galactic bulge which is stationary (on average).

Perhaps the most interesting aspect of this microlensing parallax observation is that it allows us to learn much more about this event than we can learn about ordinary microlensing events. Substituting $v_t = \tilde{v}(1-x)$ into eq. (2), we obtain

$$M(x) = \frac{1-x}{x} \frac{\tilde{v}^2 \,\hat{t}^2\, c^2}{16GL} \ . \qquad (6)$$

Fig. 2 shows $M(x)$ for the values given in Table 1 (fit # 1) assuming that $L = 8\,\mathrm{kpc}$. We can see from Fig. 2 that the lens could be either a brown dwarf in the Galactic bulge, an M dwarf between 2 and 6 kpc distant, or a more massive star nearby. If we assume that the star is currently burning its nuclear fuel, then we can also constrain this latter possibility because such a star would contribute additional unlensed flux to the observed light curve, since it would be within a few milli-arcseconds of the lensed star.

We have redone our parallax event fit including the possibility that there is some unlensed light superimposed upon the lensed star, an $R \approx 15.9$ clump giant. The best fit brightness of this assumed companion is $-4\pm 4\%$ of the normal brightness of the lensed star. We will take $0 \pm 4\%$ as our upper limit on the apparent brightness of the lens relative to the source. Then, assuming an $R$-band mass-luminosity relation of the form $L(M) = M^{2.6}$ (Mihalas & Binney, 1981) in the mass range $0.6\mathrm{M}_\odot < M < \mathrm{M}_\odot$, and allowing for 1.5 magnitudes of extinction, we obtain an upper limit on the mass of a main-sequence lens star, which we will use in Fig. 3 below.

Another, somewhat more general constraint on $x$ and $M$ can be obtained if we make use of our knowledge of the velocity distributions of the source and lensing objects, since the likelihood of obtaining the observed value of $\tilde{\mathbf{v}}$ is a strong function of the distance to the lens. For example, if the disk and bulge velocity dispersions were negligible relative to the galactic rotation velocity, then for disk lenses we would obtain a one-to-one relation $\tilde{v} = 220x/(1-x)\,\mathrm{km/s}$, and thus the lens distance here would be $D_{\mathrm{lens}} = 0.25L = 2.1\,\mathrm{kpc}$. In reality, the random motions of both disk and bulge stars broaden this relationship somewhat, but we can still obtain a useful constraint.

Given the observed $\tilde{\mathbf{v}}$, we obtain a likelihood function

$$L(x; \tilde{\mathbf{v}}) \propto \sqrt{x(1-x)}\, \rho_L(x)\, \tilde{v}(1-x)^3 \int f_S(\mathbf{v}_S)\, f_L((1-x)(\mathbf{v}_\odot + \tilde{\mathbf{v}}) + x\mathbf{v}_S)\, d\mathbf{v}_S, \qquad (7)$$

where $\rho_L$ is the density of lenses at distance $x$, and the integral is over combinations of source and lens velocity giving the observed $\tilde{\mathbf{v}}$. $\mathbf{v}_S$ and $\mathbf{v}_L = (1-x)(\mathbf{v}_\odot + \tilde{\mathbf{v}}) + x\mathbf{v}_S$ are the 2-D source and lens distribution functions (normalized to unity). We adopt a disk velocity dispersion of $30\,\mathrm{km/s}$ in each direction, with a flat rotation curve of $220\,\mathrm{km/s}$. We adopt a bulge velocity dispersion of $110\,\mathrm{km/s}$, and no bulge rotation. For the density profiles, we use a standard double-exponential disk, and a barred bulge as in Han & Gould (1995a). The source is assumed to reside in the bulge, and the lens distribution functions are the sum of those for disk and bulge.

The resulting likelihood function for $D_{\mathrm{lens}} = xL$ is shown as the solid curve in Fig. 3; it is comparable to Figure 6 of Han & Gould (1995a), and is quite insensitive to specific parameter choices. This shows that the lens is much more likely to belong to the disk than the bulge. The dashed curve is the likelihood function under the assumption of a main sequence lens. It is the direct result of multiplying the solid curve by the Gaussian probability that the lens brightness is consistent with the $0 \pm 4\%$ upper limit, so its lower amplitude is an indication that much of the region allowed by the phase space constraint is not consistent with the main sequence lens brightness limit.

It is interesting to note that inclusion of the constraint on the brightness of a main sequence lens changes the likelihood function significantly. The most probable distance from phase space and density considerations only is $D_{\mathrm{lens}} = xL \lesssim 2\,\mathrm{kpc}$, but this is excluded by the main sequence lens brightness constraint. There are two possible interpretations of this: either the lens is a main sequence star more than 2 kpc away, which is somewhat disfavored by the phase space constraint, or the lens is not a main sequence star. In the later case the lens could be an old white dwarf, or a neutron star, and it would be located at the distance indicated by the solid curve: $D_{\mathrm{lens}} = 1.7^{+1.1}_{-0.7}\,\mathrm{kpc}$ with a mass of $M = 1.3^{+1.3}_{-0.6}\mathrm{M}_\odot$ (from eq. (6)). A significant population of stellar remnant lenses might also help to explain the number of long timescale lensing events that have been observed (Bennett *et al.* 1995, Han & Gould 1995b). If the lens is a main sequence star, we must use the dashed curve, which includes the lens brightness constraint and implies $D_{\mathrm{lens}} = xL = 2.8^{+1.1}_{-0.6}\,\mathrm{kpc}$ and $M = 0.49^{+0.19}_{-0.23}\mathrm{M}_\odot$. The relative normalization of the dashed curve in Fig. 3 indicates that a main sequence lens is somewhat improbable at the $\sim 70\%$ confidence level.

An additional constraint we can place on the mass and location of the lens is due to the fact that, while the event has a high amplification of $A_{\mathrm{max}} = 10.4$ and the source has a relatively large radius of $\sim 15\,R_\odot$, we cannot detect deviations from the point-source approximation. A light curve fit allowing for a finite source size indicates that the angular Einstein ring radius must be more than 10.2 times larger than the source. Using eq. (6), this implies that $D_{\mathrm{lens}} < 7\,\mathrm{kpc}$, a result which is also implied by the likelihood function.

If it were possible to make parallax measurements for a large fraction of the observed microlensing events, it would be possible to resolve most of the uncertainties in the interpretation of present microlensing results. Toward the LMC, for example, there appear to be more events than can be accounted for by known stellar populations (Alcock *et al.* 1995a), but it is not known whether this is due to Machos in the halo, the disk, or in the LMC. Microlensing parallaxes would yield distance estimates, such as those presented here, which would be able to resolve this question for most events. Similarly, the microlensing optical depth toward the Galactic bulge seems to be higher than predicted (Udalski *et al.* 1994, Alcock *et al.* 1995b, Bennett *et al.* 1995). Suggested explanations of the high optical depth include a massive or "maximal" disk, a massive galactic bar pointed nearly along the line of sight (Paczyński *et al.* 1994, Zhou, Rich & Spergel 1995), or a substantial fraction of the lensed sources residing behind the disk (Evans 1994). While the error bars on the position determinations from individual parallax events will not always allow us to distinguish between some of these different possibilities, parallax measurements for a substantial number of events will allow us to determine the location and mass function of the bulk of the lenses.

Unfortunately, the prospects are poor that parallax measurements can be obtained for many other microlensing events from the ground. The event described here is the longest of about 50 microlensing events that have been observed so far. In addition, its peak amplification occurs close to the middle of the bulge observing season. In the same data set (Bennett *et al.* 1995), we also have two events with $\hat{t} \simeq 150$ days but with the peak amplification occurring near the ends of our observing season, so any asymmetry is weakly constrained. We have also done parallax fits for these events, but obtain results with, at best, only marginal statistical significance even though the best fit projected transverse velocities are smaller than 50 km/s. This situation could be improved somewhat in the future by obtaining more accurate photometry as events are discovered in progress (Alcock *et al.* 1994, 1995c; Udalski *et al.* 1995), but it is not possible to measure ground-based parallaxes for most of our events which have $\hat{t}$ values between 10 and 50 days.† This means that ground-based parallax measurements will not allow us to resolve the unanswered questions about the properties of the bulk of the lensing objects. However, parallaxes for almost all events could be measured by a small satellite in solar orbit, as suggested by Gould (1994b, 1995).

To summarize, we have presented the first detection of parallax in a gravitational microlensing event.‡ We have used the fit parallax event parameters along with modest assumptions about the phase space distribution of lensing objects in a likelihood analysis to estimate the distance and mass of the lens. If the lens is near the distance indicated peak of the likelihood function, then it cannot be a main sequence star because we have not seen much light from the lens. We conclude that the lens is probably either a stellar remnant or located further away than about 2 kpc, which is consistent with, but slightly disfavored by, the likelihood analysis. Finally, we've considered the possibility of obtaining parallax events for the majority of events and concluded that this requires an additional telescope in space.

## ACKNOWLEDGEMENTS

Work performed at LLNL is supported by the DOE under contract W7405-ENG-48. Work performed by the Center for Particle Astrophysics on the UC campuses is supported in part by the Office of Science and Technology Centers of NSF under cooperative agreement AST-8809616. Support from the Bilateral Science and Technology Program of the Australian Department of Industry, Science, and Technology is gratefully acknowledged. KG acknowledges DOE OJI, Sloan, and Cottrell Scholar awards. CWS thanks the Sloan and Packard Foundations for their generous support. WJS acknowledges use of Starlink computing resources at Oxford.

**References**

---

† Holz and Wald (1995) have recently shown that sufficient numbers of photons can be detected from the ground to enable parallax measurements. However, they have not addressed the systematic crowded field photometric errors, which must be improved by almost four orders of magnitude from the current state of the art in order to make ground-based microlensing parallax measurements possible.

‡ Miyamoto and Yoshii (1995) have recently claimed that our first published microlensing event (Alcock, *et al.*, 1993) shows evidence of the parallax effect, but our analysis, which is based upon twice as much data, indicates that this is not the case.

Zhao, H., Spergel, D., & Rich, M., 1995, *ApJ*, **440**, L13.

# FIGURE CAPTIONS

1. The observed two color light curve is shown as $\pm 1\sigma$ error bars, in linear units normalized to the fit unlensed brightness. The upper panel is the MACHO-R band data and the lower panel shows the MACHO-B band data. The dashed curve shows the constant velocity microlensing fit, and the solid curve shows the best-fit light curve allowing for the orbit of the Earth. Time is in days from JD 2449000.
2. The lens mass is plotted as a function of the lens distance using the $v_t$ value determined by the light curve fit shown in Fig. 1.
3. Likelihood functions for the lens distance. The solid curve indicates the likelihood function incorporating the projected velocity and the Galactic model, while the dashed curve also includes the upper limit on the brightness of a main-sequence lens.